\documentclass[twocolumn]{aastex63}

\usepackage{lineno} 
\usepackage{threeparttable}  
\usepackage{booktabs} 
\usepackage{tabularx}
\usepackage{amsmath}

\newcommand{\dsct}{$\delta$~Sct\,}

\received{2022 October 14}
\revised{2022 November 1}
\accepted{2022 November 2}

\submitjournal{ApJL}

\shorttitle{Period Luminosity Relation of Delta Scuti}
\shortauthors{Mart\'{i}nez-V\'{a}zquez et al.}

\graphicspath{{./}{figures/}}

\begin{document}

\title{A segmented period-luminosity relation for nearby extragalactic $\delta$ Scuti stars}

\correspondingauthor{C. E. Mart\'{i}nez-V\'{a}zquez}
\email{clara.martinez@noirlab.com}

\author[0000-0002-9144-7726]{C.~E.~Mart{\'\i}nez-V\'azquez}
\affiliation{Gemini Observatory/NSF's NOIRLab, 670 N. A'ohoku Place, Hilo, HI 96720, USA}
\affiliation{Cerro Tololo Inter-American Observatory/NSF's NOIRLab, Casilla 603, La Serena, Chile}

\author[0000-0002-1206-1930]{R.~Salinas}
\affiliation{Gemini Observatory/NSF's NOIRLab, Casilla 603, La Serena, Chile}

\author[0000-0003-4341-6172]{A.~K. Vivas}
\affiliation{Cerro Tololo Inter-American Observatory/NSF's NOIRLab, Casilla 603, La Serena, Chile}

\author[0000-0001-6003-8877]{M. Catelan}
\affiliation{Pontificia Universidad Cat\'olica de Chile, Facultad de F\'isica, Instituto de Astrof\'isica, Av. Vicu\~na Mackenna 4860, 782-0436 Macul, Santiago, Chile}
\affiliation{Millennium Institute of Astrophysics, Santiago, Chile}

\begin{abstract}
The period-luminosity relations (PLR) of Milky Way $\delta$ Scuti ($\delta$~Sct) stars have been described to the present day by a linear relation. However, when studying extragalactic systems such as the Magellanic Clouds and several dwarf galaxies, we notice for the first time a non-linear behaviour in the PLR of \dsct stars. Using the largest sample of $\sim$ 3700 extragalactic \dsct stars from data available in the literature -- mainly based on OGLE and SuperMACHO survey in the Large Magellanic Cloud (LMC) -- we obtain that the best fit to the period-luminosity ($M_V$) plane is given by the following piecewise linear relation with a break at $\log{P} = -1.03 \pm 0.01$ (or $0.093 \pm 0.002$~d) for shorter periods (\textit{sp}) and longer periods (\textit{lp}) than the break-point:
$$M_V^{sp} = -7.08 (\pm 0.25) \log{P} -5.74 (\pm 0.29) ;\hspace{5pt} \log{P} < -1.03$$
$$M_V^{lp} = M_V^{sp} + 4.38 (\pm 0.32) \cdot (\log{P} + 1.03 (\pm 0.01));\hspace{5pt} \log{P} \geq -1.03$$
Geometric or depth effects in the LMC, metallicity dependence, or different pulsation modes are discarded as possible causes of this segmented PLR seen in extragalactic \dsct stars. The origin of the segmented relation at $\sim$0.09 days remains unexplained based on the current data.

\end{abstract}


\section{Introduction} \label{sec:intro}
The period-luminosity relations (PLRs) of pulsating variable stars are undoubtedly one of the cornerstones of modern astrophysics. In particular, the PLR for classical Cepheids \citep{leavitt12} permitted establishing the nature of \textit{nebulae} as galaxies \citep{hubble25}, the discovery of the expansion of the Universe \citep{hubble29}, and the first rung in the distance scale, and the existence of dark energy \citep[e.g.][]{riess09}. Not as tight as the PLR for Cepheids in the visible, but uniquely tracing old ($\geq$10 Gyr) populations, the PLR of RR Lyrae stars \citep[e.g.,][]{catelan04} has also provided precise distances to, for example, the Galactic center \citep{dekany13}, globular clusters \citep[][]{Braga2015, Neeley2015, Braga2018} and dwarf galaxies in the Local Group \citep[e.g.][]{MartinezVazquez2015, MartinezVazquez2017, clara19, vivas16, Vivas2022}.

Being fainter and with shorter periods (0.008--0.42~d) and lower amplitudes (0.001--1.7 mag in $V$) than Cepheids and RR Lyrae \citep[e.g.][]{catelan15}, $\delta$ Scuti stars (hereafter $\delta$~Sct), in the intersection of the main sequence and the instability strip, are standard candles that follow not very popular PLRs. Despite their lower brightness, their value comes from tracing young, and especially \textit{intermediate} age populations, partly filling the age gap between Cepheids and RR Lyrae. They can also be more numerous than RR Lyrae stars \citep[e.g.,][]{vivas13}.

As intrinsically fainter stars, precise distances are more difficult to obtain, but also the construction of its PLR requires the determination of their pulsation mode, and to understand the effect of metallicity and evolutionary stage. Despite these difficulties, many PLRs have been explored, both theoretically and empirically \citep[e.g.][]{nemec94,mcnamara11,fiorentino15,ziaali19,jayasinghe20,barac22}. Of particular interest is the work of \citet{cohen12}, connecting both \dsct and SX~Phe stars, their metal-poor counterparts, into a single PLR, suggesting largely an independence of metallicity effects. 

In this \textit{Letter}, we revisit the \dsct PLRs given in the literature and derive a new PLR for nearby extragalactic \dsct stars. 

\begin{table}[!h]
    \centering
    \caption{Distance moduli, extinctions and metallicities.}
    \label{tab:distance}
    \setlength{\tabcolsep}{11pt}
    \begin{tabularx}{\columnwidth}{lccc}
    \toprule
    System  & $\mu_0^{(a)}$ & A$_V$ & $\langle$[Fe/H]$\rangle^{(b)}$\\
    \midrule
 LMC$^{(c)}$    & 18.476   & 0.40   & $-$0.5 \\
 NGC~1846       & 18.45    & 0.09   & $-$0.5 \\
 SMC            & 18.97    & 0.28   & $-$1.0 \\
 NGC~419        & 18.85    & 0.15   & $-$0.55\\ 
 Carina         & 20.01    & 0.09   & $-$1.0 \\
 Fornax         & 20.72    & 0.06   & $-$1.7 \\
 Sculptor       & 19.62    & 0.06   & $-$1.7 \\
 Sextans        & 19.64    & 0.14   & $-$1.9 \\
 GGCs           &   :      &   :    &  [$-$2.35, $-$0.59] \\       
Galactic Field  & $^{(d)}$ &  $^{(d)}$ & [$-$0.5, 0.3]  \\     
    \bottomrule
    \end{tabularx}
    \begin{scriptsize}
    \begin{tablenotes}
    \item \textit{(a)} References for distances and average extinction: 
    LMC \citep{pietrzynski19, westerlund97}; 
    NGC~1846 \citep{goudfrooij09}; 
    SMC \citep{graczyk14, westerlund97}; 
    NGC~419 \citep{goudfrooij14};
    Carina \citep{coppola15};
    Fornax  \citep{rizzi07}; 
    Sculptor \citep{martinezvazquez16};
    Sextans \citep{vivas19};
    GGCs  (\citealt{baumgardt18}, \citealt[][2010 edition]{harris96}).    
    \item \textit{(b)} References for mean metallicities: LMC and NGC~1846 \citep{grocholski06, carrera08a}; SMC and NGC~419 \citep[][A. Mucciarelli priv. comm.]{carrera08b,mucciarelli14}; Sextans \citep{kirby11}; Carina \citep{koch06}; Fornax and Sculptor \citep{kirby13}; GGCs \citep{carretta09};  Galactic Field \citep{jayasinghe20}.
    \item \textit{(c)} For the LMC we use individual instead of the mean distance and absorption values. See \S~\ref{sec:lmc} for more details.
    \item \textit{(d)} For the Galactic Field we used the M$_V$ values provided by \citet{ziaali19}, thus $\mu_0$ and A$_V$ were not required.
    \end{tablenotes}
    \end{scriptsize}
\end{table}

\section{Data} \label{sec:data}

We have gathered data available in the literature associated with \dsct and SX~Phe studies to perform a comprehensive analysis of their PLR. To do that, we have used several catalogs that report periods and mean magnitudes (particularly in the $V$-band) for the \dsct and SX~Phe stars (collectively also known as Dwarf Cepheids; see \citealt{mateo93, vivas13}).

For the Large Magellanic Cloud (LMC) we have used the \dsct stars catalogs of OGLE-III \citep{poleski10} and SuperMACHO \citep{garg2010}. Additionally, we have included the \dsct from the LMC cluster NGC~1846 and its surrounding field \citep{salinas18}. For the Small Magellanic Cloud (SMC), we have used OGLE-II \citep{soszynski02} and the recent work of \citet{martinezvazquez21a} in the SMC cluster NGC~419. 

For the \dsct/SX~Phe stars in dwarf galaxies in the Local Group we have collected the data from the following works: Fornax \citep{poretti08}, Carina \citep{vivas13, coppola15}, Sculptor \citep{martinezvazquez16}, and Sextans \citep{vivas19}.

As a comparison to the extragalactic sample, we have used different Galactic \dsct studies. To start, we have employed the \citet{ziaali19} catalog, which is a compilation of the \citet{rodriguez2000} catalog; and in addition, \dsct stars studied with {\em Kepler} \citep{murphy19}, with distances from \textit{Gaia} DR2 \citep{gaiadr2}. Finally, for the Galactic globular clusters we have employed the compilation made by \citet{cohen12}. 

Table~\ref{tab:distance} lists the most updated distance moduli available in the literature and mean extinctions used in this work for each of these systems.

\begin{figure}[!h]
    \centering
    \hspace{-0.7cm}
    \includegraphics[width=0.45\textwidth]{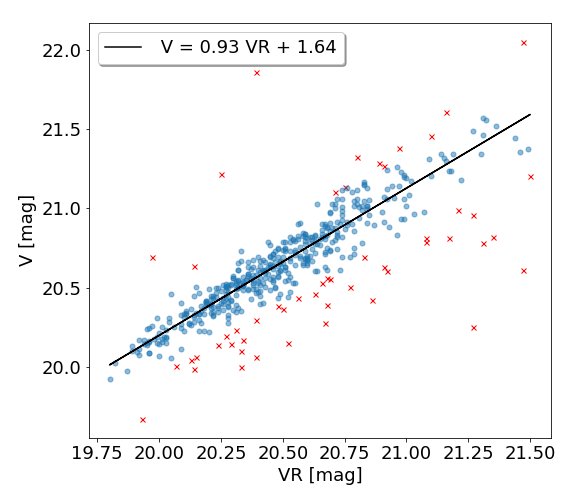}
    \caption{$VR$ versus $V$ for the stars in common between OGLE and SuperMACHO. The black line shows the fit to the data.  Only stars represented with blue filled circles have been used in the fit, while stars shown as open red crosses have been rejected by an iterative 2.5$\sigma$-clipping algorithm.}
    \label{fig:V_VR_relation}
\end{figure}

\section{Period Luminosity relation of \dsct stars} \label{sec:pl}

In the following, we will show how a break in the PLR of extragalactic \dsct stars arises from literature data, and how it is largely independent on the assumptions used for distance, reddening, and metallicity effects.

\subsection{LMC}\label{sec:lmc}

The OGLE and SuperMACHO survey discovered more than $\sim$4500 \dsct stars in the LMC \citep[][ respectively]{poleski10,garg2010}. However, due to the different goals of both surveys, the period and magnitude distributions obtained were significantly different, being longer and brighter in OGLE, and shorter and fainter in SuperMACHO, with $\sim$500 \dsct stars in common. Because SuperMACHO used a broad $VR$ filter instead of $V$, we use these stars in common between the two catalogs in order to obtain the transformation to the $V$-band. The transformation between $VR$ and $V$ did not show any color-term dependency for these stars. Figure~\ref{fig:V_VR_relation} shows the graphic representation of the relation between $VR$ (SuperMACHO) and $V$ (OGLE). After applying a 2.5 sigma-clipping over the $V$ versus $VR$ fit, we obtained the following relation:

\begin{equation}
V = 0.93\, (\pm0.01) \cdot VR + 1.6\, (\pm0.3) .
\end{equation}

The black line in Figure~\ref{fig:V_VR_relation} is the fit to the data (blue circles) and the red crosses are those stars rejected from the $\sigma$-clipping. The Pearson coefficient ($r = 0.95$) and the RMS (0.09 mag) of this relation is also an indication that the transformation between these two bands for these stars follows a linear dependence and does not depend on other factors.

\begin{figure}
    \centering
    \hspace{-0.7cm}
    \includegraphics[width=0.5\textwidth]{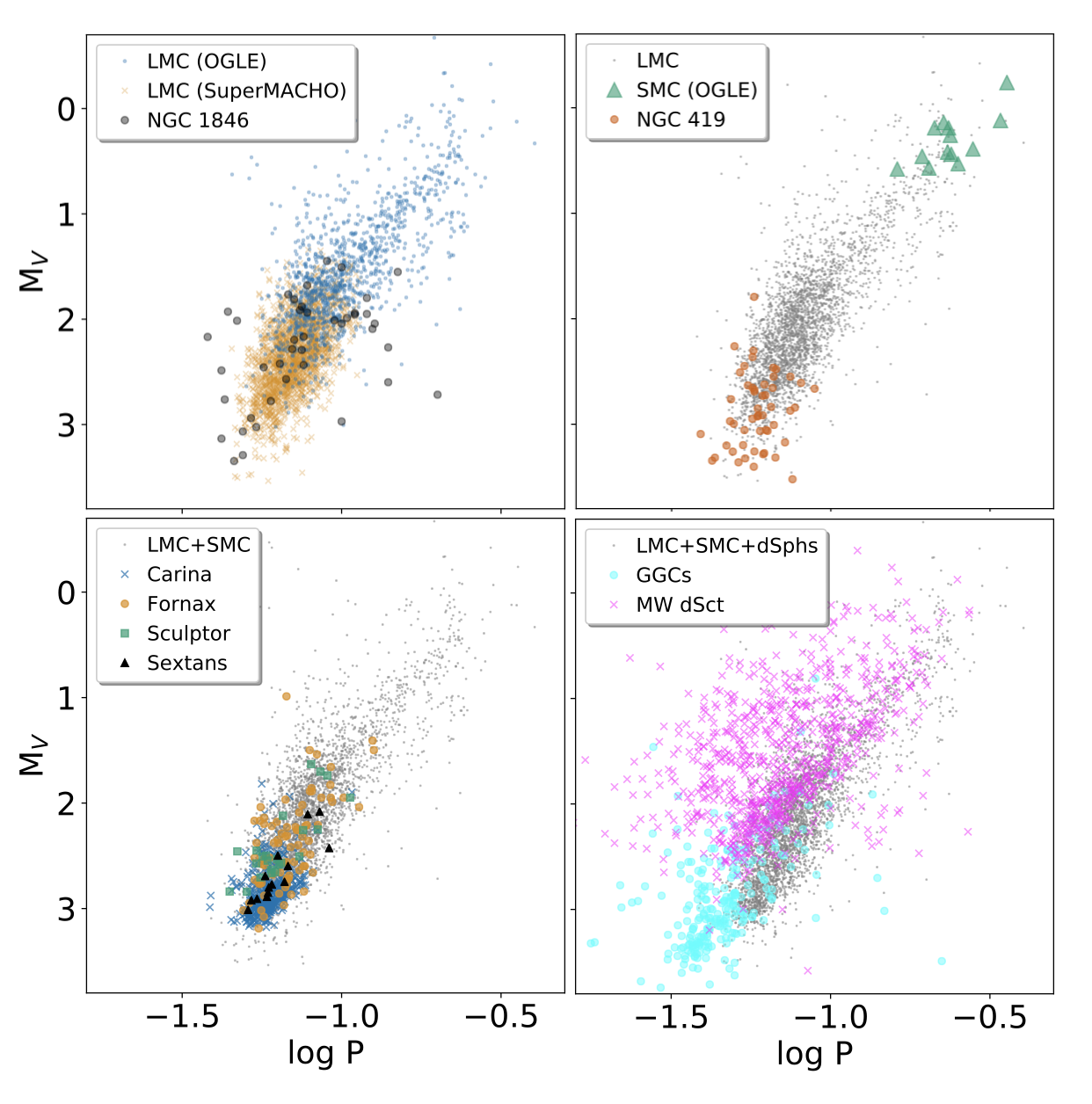}
    \caption{Period versus absolute magnitude in $V$ plane of \dsct stars in different environments.
    \textit{Top left.} The LMC. 
    \textit{Top right.} The SMC. 
    For guidance, background grey dots represent the \dsct stars from the LMC (combined OGLE + SuperMACHO sample).
    \textit{Bottom left.} dSph galaxies. 
    For guidance, background grey dots represent the \dsct stars from the LMC and the SMC.
    \textit{Bottom right.} Galactic (field and globular clusters) \dsct stars. Background grey dots are the combined \dsct samples from the LMC, the SMC, and dSph galaxies. 
    See main text for references for each sample.}
    \label{fig:pl_dsct_all}
\end{figure}

The top left panel of Figure~\ref{fig:pl_dsct_all} shows the PLR in $M_V$ for the LMC \dsct stars. In order to obtain a clean sample of \dsct stars in the LMC, we remove those stars that are flagged with a remark in the \citet{poleski10} catalog. The vast majority of those removed (1488) are uncertain \dsct stars, while 19 are Galactic \dsct stars based on their proper motions and 5 with ``variable mean luminosities''. On the other side, for the \citet{garg2010} \dsct star catalog we did not apply any restriction. 

We detect an interesting feature to our knowledge never reported in previous studies; the PLR of the \dsct in the LMC follows a broken power law rather than a single linear relation over the full period range ($-1.4 < \log{P} < -0.4$). Particularly, a segmented linear relation with a break-point at $\log$ $P$ $\simeq$ -1.03 reproduces better this behaviour (see \S~\ref{sec:pl_fit}). 

It is worth noting that absolute $V$ magnitudes were corrected after applying the proper A$_V$ extinction to each individual star, using the LMC reddening map from \citet{haschke11}. In addition, because of the extended structure of the LMC in the sky and its proximity, we correct the individual distances by the LMC geometry \citep{vdm01}. Finally, we explored the influence of the depth of the LMC by generating random samples using the measured depths in the different parts of the LMC given by \citet{subramanian09}. We conclude that these effects are negligible, and that they are not causing a change of slope in the PLR. The top panel of Figure~\ref{fig:pl_dsct_all} includes the corrections by reddening and the LMC geometry.

Additionally, \citet{salinas18} found 55 \dsct stars in a 5.5\arcmin$\times$5.5\arcmin\, area centered on the LMC globular cluster NGC~1846. Over 40 of them are outside 2 half-light radii, and probably belong to the LMC field. The \dsct stars of \citet{salinas18} are represented by black circles in the top left panel of Figure~\ref{fig:pl_dsct_all}\footnote{We excluded in this plot the \dsct stars in NGC~1846 with probable aliased periods according to \citet{salinas18}.}. The majority of these stars are located around the bulk of the SuperMACHO sample, and except for a few outliers, i.e., around the region of shorter periods, following a similar tendency.

\subsection{SMC}\label{sec:smc}
The only search for \dsct stars in the SMC before 2021 was done by OGLE. \citet{soszynski02} detected 19 candidates that they classified as ``Other variable stars'' but they claimed that most of them are probably \dsct stars. From this list, 17 have a period compatible with being \dsct stars (P $\la$ 0.3 d)\footnote{OGLE004616.17-731416.1 and OGLE005507.46-724434.0 have periods of 0.50 and 0.57 days, respectively.}. By visual inspection of the individual light curves, using periods reported by OGLE, two of them do not show a reliable light curve\footnote{OGLE005008.48-725916.5 and OGLE005259.56-725605.3}. Therefore, we end up with 15 stars detected by OGLE in the SMC as probable \dsct stars.

In a recent work made by \citet{martinezvazquez21a}, we have detected 54 \dsct stars in a field of 5.5\arcmin$\times$5.5\arcmin~centered on the SMC globular cluster NGC~419. A total of 48 \dsct stars were two half-light radii outside of NGC~419, and therefore consistent with being \dsct stars of the SMC field. 

The top right panel of Figure~\ref{fig:pl_dsct_all} shows the PLR of the \dsct stars detected in the field of the SMC so far. Green triangles are the 15 \dsct stars detected by \citet{soszynski02}, while the orange circles represent those \dsct stars discovered by \citet{martinezvazquez21a} in the field of the SMC globular cluster NGC~419. For clarity, a few \dsct stars that seem to have aliased periods in the latter work were removed. As a guidance, we plot the LMC \dsct stars that were displayed in the top left panel of Figure~\ref{fig:pl_dsct_all} as background grey dots. The difference between the period range (and luminosity) of both works is understandable given the nature of each study, and it reflects the need of obtaining a more complete study of \dsct stars in the SMC. While the observation strategy of OGLE is better suited to detect variables like RR Lyrae and Cepheids that have timescales for variability $> 0.5$~d (rather than a few hours) --detecting therefore only the \dsct stars with longer periods--, the strategy of \citet{martinezvazquez21a} was focused on the study of \dsct stars. Besides, \dsct stars were near the faint limit of OGLE and the most affected stars in terms of detection are the shorter period \dsct stars since they are fainter. It is also worth noting that because of the time span of the observations of NGC~419, the \dsct stars with periods larger than 0.2 days were difficult to detect in this study.

It is noticeable by looking at this panel how the larger and the shorter period \dsct stars in the SMC follow the same tendency as in the LMC. This also supports the argument that the PLR of the \dsct stars cannot be explained by only one linear relation but a segmented linear relation with a break-point at $\log$ $P$ $\simeq$ -1.03.

\subsection{Dwarf spheroidal galaxies}\label{sec:dSphs}
It was not until the past decade that \dsct stars in dwarf galaxies started becoming relevant. Several factors played a fundamental role in this kind of study, the most important being deep photometry and high cadence. \citet{mateo98} discovered 20 \dsct stars in Carina. More than fifteen years later, \citet{vivas13} reported the discovery of 340 \dsct stars in Carina and \citet{coppola15} increased the sample of \dsct stars in Carina in over 100 more. So far, Carina is the dwarf spheroidal (dSph) galaxy with the largest sample of \dsct stars known, 426 \dsct stars from the merger of the two previous works \citep{vivas19}. The second largest catalog in terms of \dsct stars in dSph galaxies is Fornax with 85 \dsct stars detected by \citet{poretti08}, followed by Sculptor with 23 \dsct \citep{martinezvazquez16} and Sextans with 14 \dsct stars \citep{vivas19}. 

Carina and Fornax have extended star-formation histories \citep[see, e.g., ][]{gallart15} and therefore they will probably have a mixture of \dsct stars coming from an old and metal-poor population (SX~Phe) and \dsct stars coming from a younger and more metal-rich population. In Sculptor and Sextans, on the other hand, all their sample is composed by an old and metal-poor population, since both galaxies had an event of star formation more than 10 Gyr ago \citep{deBoer2012, Bettinelli2018}.

The bottom left panel of Figure~\ref{fig:pl_dsct_all} shows the PL plane of the \dsct stars detected in each dSph. Again, there is a good agreement with what we have seen in previous panels for the shorter periods of \dsct stars.

\subsection{Milky Way: field stars and globular clusters}
Finally, we display in the bottom right panel of Figure~\ref{fig:pl_dsct_all} the Galactic \dsct stars and the SX~Phe that come from the Galactic globular clusters (GGCs). Interestingly, we can notice here a dual behaviour that may be due to the different populations traced. While the GGCs harbor an old metal-poor population of \dsct (SX~Phe) stars, the Galactic field \dsct stars are representatives of a metal-rich population. In fact, the fit that \citet{ziaali19} makes is in agreement with that from \citet{mcnamara11} for [Fe/H] = 0.0 dex (see \S~\ref{sec:comparison}). 

Another aspect to mention here is that the lower end of the global period-luminosity plane of \dsct stars is ruled by the SX~Phe from the GGCs, which do not follow the same relation than the MW field \dsct stars \citep[see,][]{fiorentino15}.

As mentioned in \S~\ref{sec:data}, we used the Galactic \dsct sample as comparison but we won't be using these data in the derivation of the PLR.

\subsection{A piecewise linear relation for the \dsct's PLR}\label{sec:pl_fit}

We compiled all the extragalactic data mentioned in the previous subsections and using an orthogonal distance regression \citep{BOGGS1988169} we fit a piecewise linear relation to these data. To make the fit less sensitive to outliers, we used only those stars falling in regions of the period-$M_V$ plane where the relative density of \dsct stars is at least 10\%. The slopes and the break-point of the piecewise linear relation are as follow:

\begin{figure}
    \centering
    \hspace{-0.7cm}
    \includegraphics[width=0.5\textwidth]{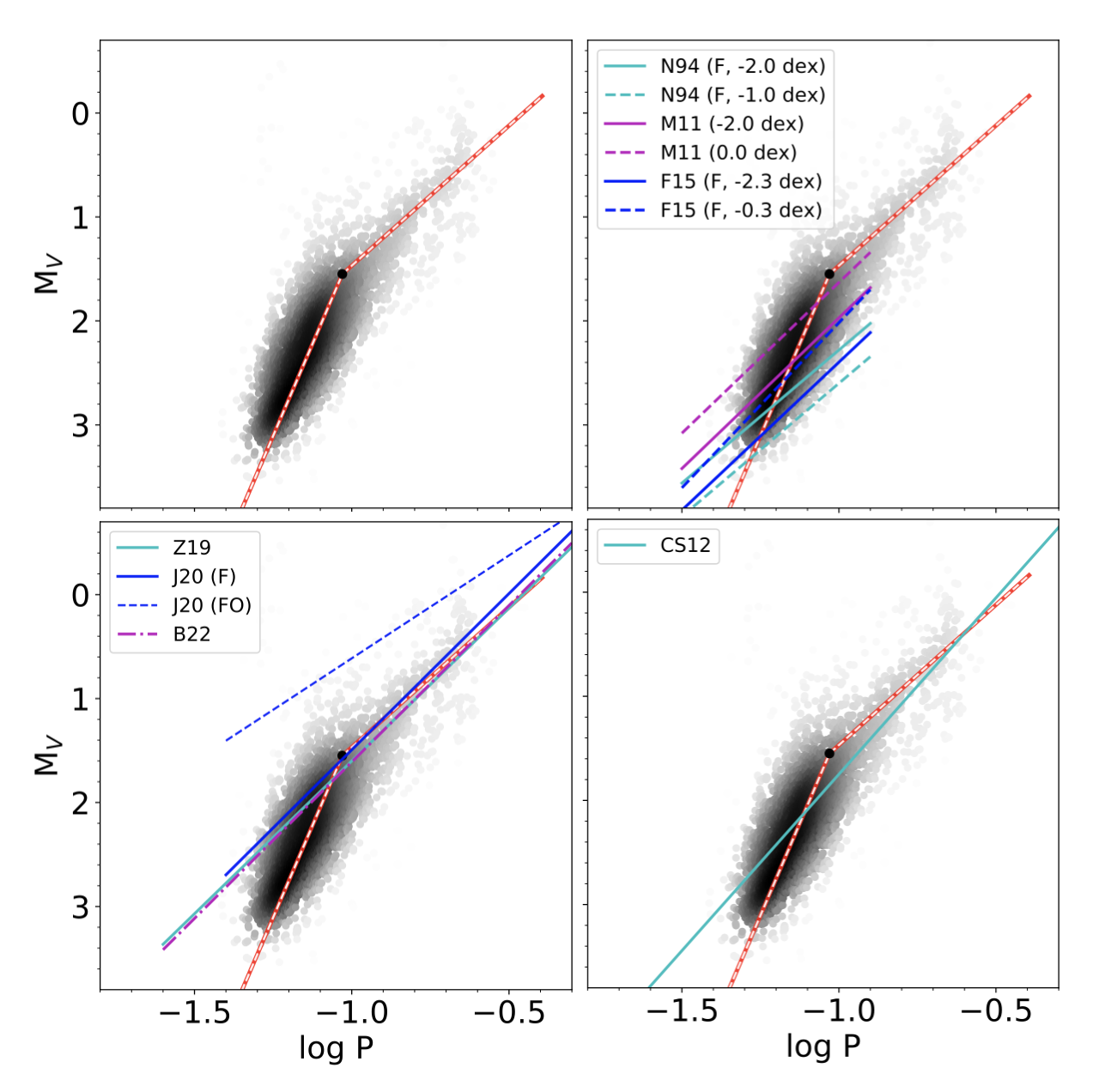}
    \caption{Density map of the \dsct stars in the period versus absolute magnitude in $V$ plane. The top left panel shows the fit obtained in this work while the rest of the panels show a comparison between this new fit and previous PLR available in the literature. \textit{References}: N94 \citep{nemec94}, M11 \citep{mcnamara11}, CS12 \citep{cohen12}, F15 \citep{fiorentino15}, Z19 \citep{ziaali19}, J20 \citep{jayasinghe20}.}
    \label{fig:pl_fit_all}
\end{figure}

\begin{small}
\begin{equation}\tag{2a}\label{eq:pl_sh}
M_V^{sp} = -7.08 (\pm 0.25) \log{P} -5.74 (\pm 0.29)
\\ ;\hspace{5pt} \log{P} < -1.03
\end{equation}
\vspace{-0.5cm}
\begin{equation}\tag{2b}\label{eq:pl_lg}
M_V^{lp} = M_V^{sp} + 4.38 (\pm 0.32) (\log{P} + 1.03 (\pm 0.01)) 
\\ ;\hspace{5pt} \log{P} \geq -1.03
\end{equation}
\end{small}

\noindent where \textit{sp} and \textit{lp} superscripts stand for \textit{shorter periods} and \textit{longer periods} than the break-point, respectively. The break-point derived is $\log{P} = -1.03 \pm 0.01$ ($P = 0.093 \pm 0.002$ days). The goodness of the fit given by the residual variance ($S_r^2= 0.003$) indicates that this piecewise linear relation represents better these data than a single linear relation ($S_r^2= 0.02$).

The newly derived PLR for extragalactic \dsct stars is shown in the top left panel of Figure~\ref{fig:pl_fit_all}. 

\subsection{Comparison with previous relations}\label{sec:comparison}

During the past few decades, there have been several efforts to obtain a PLR for the \dsct stars. Some teams (\citealt{nemec94} and \citealt{fiorentino15}) derived relations for different \dsct pulsators (especially for fundamental, F, and first overtone, FO), but others \citep{mcnamara11} obtained the PLR for only F \dsct. In practice, it is very difficult to distinguish different pulsation modes from an observational point of view \citep[see e.g.,][]{soszynski02}. This is the reason why we did not make any distinction in \S~\ref{sec:pl_fit} when obtaining our PLR  (Eqs.~\ref{eq:pl_lg} and \ref{eq:pl_sh}). Moreover, OGLE and SuperMACHO first overtone and double-mode pulsators, when present, occupy basically the same position in the period-M$_V$ plane than the fundamental pulsators.

Figure~\ref{fig:pl_fit_all} shows a comparison between our derived PLR and previously obtained relations. Particularly, the top right panel displays the PL obtained by \citet{nemec94,mcnamara11,fiorentino15}. The pioneering work of \citet{nemec94} assessed a PLR using 21 (15 F and 6 FO) \dsct (or SX~Phe) stars from GGCs, while \citet{mcnamara11} used 20 \dsct stars of the field of the Milky Way. On the other hand, the \citet{fiorentino15} relationships shown here is purely theoretical, obtained from pulsation models for main-sequence stars computed assuming $Z=0.0001$ and $Z=0.008$ (i.e., [Fe/H] = $-2.3$ and $-0.3$, since they consider $\alpha$-enhanced stellar evolution models). 

The bottom left panel of Figure~\ref{fig:pl_fit_all} shows the comparison with the recently derived PLRs for the Galactic \dsct stars using different data sets. \citet{ziaali19} is based on 228 stars from \citet{rodriguez2000} catalog plus 1124 stars observed by the \textit{Kepler} mission, while the \citet{jayasinghe20} catalog used the all-sky catalog of $\sim$8400 \dsct stars in ASAS-SN, both using distances from \textit{Gaia} DR2. The \citet{ziaali19} relation is very similar to the \citet{mcnamara11} relation but the former was extended in the longer period range that is seen only in the Galactic and Magellanic Clouds \dsct stars. Both relations are similar (despite an offset in the zero-point) and agree with the long-period end of the piecewise PLR obtained in this work. More recently, \citet{barac22} revised \citet{ziaali19} PLR using Gaia DR3 parallaxes also obtaining similar results.

The bottom right panel of Figure~\ref{fig:pl_fit_all} presents the comparison with the relation obtained by \citet{cohen12}, who made a fit over the entire period range of \dsct stars (including the SX~Phe coming from the GGCs). There is a clear discrepancy between both relations. However, the \citet{cohen12} relation was obtained using only some of the \citet{poleski10} \dsct stars in the LMC, the very few \dsct stars that were known in Carina \citep{mateo98}, and none from Sculptor and Sextans, being Fornax \citep{poretti08} the most significant one in terms of \dsct stars at that time. The inclusion of stars discovered after 2012 in the dSph galaxies mentioned (Carina: \citealt{vivas13,coppola15}, Sculptor: \citealt{martinezvazquez16}, Sextans:\citealt{vivas19}), plus the LMC \dsct stars detected by \citet{garg2010}, in the period-M$_V$ plane opens a new debate about the origin of the PLR in the \dsct stars since it is clearly noticeable that a new behaviour is governing the shorter period range. 

\subsection{Possible causes}\label{sec:effects}

In the following we will review the main causes that could explain this behaviour in the PLR of the \dsct stars.

\begin{itemize}
    \item \textbf{Metallicity effect}. The fact that the break in the PLR of the \dsct stars comes from the metallicity is unlikely. The strongest argument that supports this is that the main contributions to Figure~\ref{fig:pl_fit_all} come from \citet{poleski10} and \citet{garg2010}, and both samples cover the same regions of the LMC. However, they obey different relations. While the \citet{poleski10} sample represents larger period range and follows eq.~\ref{eq:pl_lg}, the \citet{garg2010} sample fills the lower period end and follows better eq.~\ref{eq:pl_sh}. The existence of any metallicity gradients should be present in both samples (i.e., for longer and shorter periods), 
    and not be reflected in only one of them. In addition, bearing in mind that the metallicity of the LMC is between [Fe/H] = --0.3 and --0.7 dex \citep{grocholski06,carrera08}, there are no PLR that can explain, with these metallicities, the large luminosity dispersion observed in the short period range (see Figure~\ref{fig:pl_fit_all}). Although we do not discard the fact that there may be metallicity dependence in the PL, we point out that this would not be the main driver that is causing this in the PLR of the \dsct stars.
    
    Moreover, by looking at Figure~\ref{fig:pl_dsct_all} (especially the bottom left panel) we see that the inclusion of several systems with different metallicities (see Table~\ref{tab:distance}) does not produce any clear trend in the luminosity-period relation predicted by the metallicity of previously derived relations (see \S~\ref{sec:comparison}, top right panel of Figure~\ref{fig:pl_dsct_all}). Those systems seem to follow the same relation.  
    
    \item \textbf{Depth and geometry effects.} This effect was shown to be almost negligible in the LMC (see \S~\ref{sec:lmc}) and it would be less significant for farther galaxies. 

    \item \textbf{Pulsation mode.} We also investigated the possible association of this broken relation as due to the pulsation mode of the \dsct stars. The different pulsators (fundamental, first overtone, multi-mode) in the \citet{poleski10} and \citet{garg2010} samples did not shed any light on this hypothesis. First overtone and multi-mode pulsators are \emph{brighter} than fundamental pulsators \citep[see e.g.,][]{jayasinghe20}, therefore they cannot be the reason of the behaviour seen at fainter and shorter periods. Thus, we also discarded the pulsation mode as a possible origin of the broken PLR relation.
\end{itemize}

We speculate that a segmented PLR may be naturally present in sufficiently metal-poor systems (i.e., with LMC-like metallicities or lower), and that an attempt to reproduce this behavior should be made using state-of-the-art stellar pulsation and evolution models, coupled with synthetic populations tools that properly mimic the properties of \dsct and SX~Phe stars in the metal-poor systems studied in this paper.

A similar behaviour (although less pronounced) has been observed by in the PLR of Cepheids of the LMC \citep[e.g.][]{Bhardwaj2020}. They could not explain either the physics behind it. The broken power-law relation seen in the \dsct stars is evident in the data collected from the extragalactic sources and future investigations will provide more knowledge about the physics behind this particular behaviour.

\section{Final remarks}\label{sec:conclusions}
In this \textit{Letter}, we have used 3664 \dsct stars from 8 different extragalactic stellar systems. Such a large sample, spanning all period range for this type of stars, has revealed a broken PLR not seen before in the MW \dsct stars. In particular, we statistically tested that the best representation of the extragalactic \dsct stars is by a piecewise linear relation of the form given by Eqs.~(\ref{eq:pl_sh}) and (\ref{eq:pl_lg}).  

We have shown that this new dependence is not due to depth or geometric effects in the extragalactic sources as they proved to be negligible. Also, it most likely not caused by metallicity effects, since both SuperMACHO and OGLE studied the same region of the LMC, therefore probing the same metallicity range. Furthermore, the pulsation mode of the \dsct stars seems not to be causing this effect either. The origin of the segmented relation at $\sim$0.09 days remains still unknown based on the current data. A new survey focused on short period variables, delivering a homogeneous sample of \dsct, would either confirm or reject the new PLR proposed here.

\software{Scipy \citep{Scipy}, Astropy \citep{Astropy}, Numpy, \citep{Numpy}, Pandas \citep{Pandas}, Matplotlib \citep{Matplotlib}, TOPCAT \citep{TOPCAT}}

\acknowledgments

C.~E.~M-V would like to dedicate this paper to the memory of her late father. C.~E.~M-V also thanks Elham Ziaali for providing the catalog of Galactic \dsct stars and John Blakeslee for useful conversations and advice. We are also grateful to the anonymous referee for helpful suggestions.

C.~E.~M-V and R.~S. are supported by the international Gemini Observatory, a program of NSF's NOIRLab, which is managed by the Association of Universities for Research in Astronomy (AURA) under a cooperative agreement with the National Science Foundation, on behalf of the Gemini partnership of Argentina, Brazil, Canada, Chile, the Republic of Korea, and the United States of America.

\clearpage
\newpage
\bibliographystyle{aasjournal}

\clearpage

\end{document}